# Image-Based Abnormal Data Detection and Cleaning Algorithm via Wind Power Curve

Huan Long , *Member, IEEE*, Linwei Sang , Zaijun Wu , *Member, IEEE*, and Wei Gu , *Senior Member, IEEE*

*Abstract*—This paper proposes an image-based algorithm for detecting and cleaning the wind turbine abnormal data based on wind power curve (WPC) images. The abnormal data are categorized into three types, negative points, scattered points, and stacked points. The proposed algorithm includes three steps, data pre-cleaning, normal data extraction, and data marking. The negative abnormal points, whose wind speed is greater than cut-in speed and power is below zero, are first filtered in the data pre-cleaning step. The scatter figure of the rest wind power data forms the WPC image and corresponding binary image. In the normal data extraction step, the principle part of the WPC binary image, representing the normal data, is extracted by the mathematical morphology operation (MMO). The optimal parameter setting of MMO is determined by minimizing the dissimilarity between the extracted principle part and the reference WPC image based on Hu moments. In the data mark step, the pixel points of scattered and stacked abnormal data are successively identified. The mapping relationship between the wind power points and image pixel points is built to mark the wind turbine normal and abnormal data. The proposed image-based algorithm is compared with *k*-means, local outlier factor, combined algorithm based on change point grouping algorithm and quartile algorithm (CA). Numerous experiments based on 33 wind turbines from two wind farms are conducted to validate the effectiveness, efficiency, and universality of the proposed method.

## I. INTRODUCTION

THE wind turbines are rapidly installed around the world over past decades. With the maturity of the wind turbine design and construction technology, the wind farm operations and maintenance (O&M) has gradually attracted more and more attentions [1]. The wind turbine data from the Supervisory Control And Data Acquisition (SCADA) system, implying the performance of wind turbine, is widely utilized to wind turbine conditional monitoring [2], wind power prediction [3] and other applications [4]. However, the SCADA data contain different types of the abnormal data. The conventional data cleaning approaches cannot detect all of the abnormal data well, especially the stacked abnormal data. They are easily recognized as the normal data when the amount is large.

The factors causing the wind turbine abnormal are various, including wind turbine faults, wind curtailment, extreme weather conditions and so on. The wind turbine faults also have different types, such as pitch faults, lower gearbox oil, yaw problem, generator brushes worn. Different types of faults bring in different abnormal data. In China, the wind curtailment has been a serious problem to further develop the wind power with a national average curtailment rate of 17.1% in 2017 [5]. The wind curtailment command makes the anomaly curvature of the wind power curve, affecting wind turbine O&M schedule. Thus, detecting and cleaning the abnormal data of the wind turbine is crucial for wind power curve application.

Previous research works of detection and cleaning of wind power abnormal data can be classified to two categories. The first one utilizes the statistical characteristics of the abnormal data different from the normal data to clean them, including data density, distance, variance and so on [6]–[10]. Zheng *et al.* [6] divided the wind power data into six categories and utilized LOF method to detect the irrational and unnatural data. The weighted distances of the *k*-nearest neighbors were calculated to evaluate the outlier degree. Yesilbudak [8] proposed a *k*-means clustering algorithm based on squared Euclidean and City-Block distance measures to clean the abnormal data. Since the clustering methods [6], [8] are based on the data density or distance, they are suitable to detect the scattered abnormal data and easily miss the stacked abnormal data. Zhao *et al.* [9] combined the quartile method and density-based clustering method to clean the data. The quartile method first detected the scattered outliers and the density-based spatial clustering method then cleaned the stacked outliers. Shen *et al.* [7] combined the change point grouping algorithm and quartile algorithm to detect the outliers. The wind speed was divided into several bins, the variance change rates of the ordered wind power points in each bin were calculated to detect the change points which were regarded as the abnormal data. The quartile algorithm was then used to clean the rest abnormal data.

Although the refs [7], [9] considered several algorithms to filter the stacked data which were missed by the clustering algorithms, it still did not work well for the abundant stacked abnormal data. It is because the above methods assume the normal data are major and use the statistical characteristics of the major data to detect the abnormal data. When the wind curtailment frequently appears, the large abnormal data will become the major part and be regarded as the normal data.

Manuscript received January 9, 2019; revised March 29, 2019 and April 22, 2019; accepted April 25, 2019. Date of publication April 30, 2019; date of current version March 23, 2020. This work was supported in part by the National Natural Science Foundation of China under Grant 51807023 and in part by the Natural Science Foundation of Jiangsu Province under Grant BK20180382. Paper no. TSTE-00029-2019. *(Corresponding author: Huan Long.)*

The authors are with the School of Electrical Engineering, Southeast University, Nanjing 210096, China (e-mail: hlong@seu.edu.cn; sanglinwei21@163.com; zjwu@seu.edu.cn; wgu@seu.edu.cn).

Color versions of one or more of the figures in this paper are available online at http://ieeexplore.ieee.org.

Digital Object Identifier 10.1109/TSTE.2019.2914089

The second type of method models the wind power curve with the help of large normal data [11]–[13]. The data located outside the boundary of the wind power curve are regarded as the outlier points. Kusiak et al. [11] constructed the wind power curve model by the k-nearest neighborhood to detect anomalies online. The value of parameter k was needed to be set properly for each wind turbine. Ye et al. [12] modelled a probabilistic wind power curve based on the copula conditional quantile method. Wang et al. [13] proposed a Gaussian mixture copula model to fit the wind power curves. The outliers were excluded by a probability contour. However, the stacked abnormal data also could not be effectively filtered. In, the large amount normal data are needed to train a reliable wind power curve. Due to the performance of the abnormal data detection totally relying on the training data, the universality of the trained power curve is limited. The setting of the algorithm parameters is also adjusted for the different wind turbines.

In summary, the current methods to detect the abnormal data have several drawbacks: i) the stacked abnormal data cannot be filtered effectively, especially when the amount is large; ii) plenty of the normal data are needed to train the reliable power curve model; iii) the setting of the model parameters is determined case by case and the universality of the model is limited.

On the account of the above drawbacks, this paper proposes an image-based algorithm to detect and clean the abnormal data through the WPC image. The image recognition technology has been widely used in the electricity field, such as electricity meter reading [14], icing thickness detection of power transmission lines [15], electric tower detection [16] and so on.

In the proposed image-based algorithm, three steps, data pre-cleaning, normal data extraction and data marking, are included. The abnormal data can be divided into the negative abnormal data, the scattered abnormal data and stacked abnormal data. The negative abnormal points are first filtered by the pre-cleaning process. The rest data are converted into the WPC binary image. The principle part of the WPC binary image, representing the normal data, is then extracted by MMO algorithm. The dissimilarity between the extracted principle part and the reference WPC is minimized to determine the optimal MMO parameter based on Hu moments. In the data marking step, the pixel points of the scattered and stacked abnormal data are sequentially identified. The normal and abnormal data are marked based on the mapping relationship between the pixel points and actual power points. The abnormal data are more intuitive to be identified in the form of the WPC image than the traditional statistical methods or data mining methods. Besides, it is immune on the impact of abundant stacked abnormal data on the statistical characteristics. The cases of 33 wind turbines from two wind farms validate the effectiveness, efficiency and universality of the proposed image-based algorithm.

## II. Background of Wind Power Curve

The WPC is formed based on the data collected from the SCADA system, representing the relationship between the wind power and wind speed. It directly reflects the performance of the wind turbine control system. However, various factors cause the WPC abnormity, such as the wind curtailment command, wind turbine faults, extreme weather conditions and so on. The abnormal data based on WPC can be included into three types, the negative abnormal data, the scattered abnormal data and the stacked abnormal data, shown in Fig. 1.

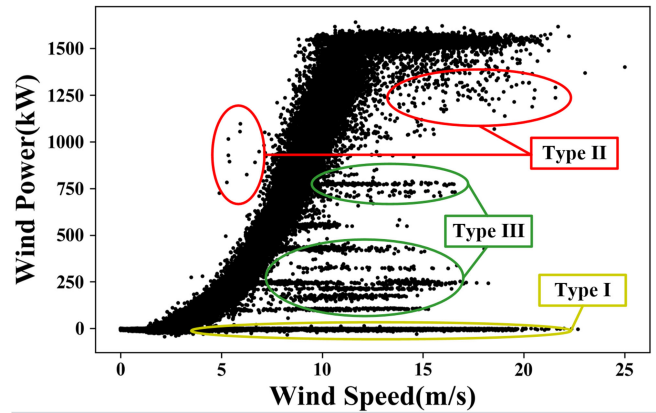

Fig. 1. Types of the abnormal data based on the wind power curve.

### A. Type I Abnormal Data

The Type I abnormal data are the negative abnormal data whose wind speed are greater than the cut-in speed and power are below zero. When the wind turbine absorbs the power from the grid, the power output will be recorded as the negative value. Most of the negative abnormal data are close to zero. The main reasons generating Type I abnormal data include the unplanned maintenance, wind turbine failure and wind curtailment.

### B. Type II Abnormal Data

The Type II abnormal data are the scattered abnormal data. They are randomly distributed around the normal curvature. The sensor fault, sensor noise and some uncontrolled random factors will cause these abnormal data. Thus, they follow the haphazard distribution and appear unpredictably and discontinuously.

### C. Type III Abnormal Data

The Type III abnormal data are the stacked abnormal data. These abnormal data usually appear in a consecutive period of time and stacked in a line in the power curve. Thus, they are easily recognized as the normal data when the amount is large. They are usually caused by the wind curtailment command or communication failures, especially the wind curtailment command. In China, the wind curtailment has become one of the major problems restricting China wind power large scale development.

## III. Image-Based Abnormal Data Detection and Cleaning Algorithm

In this section, the process of the proposed image-based algorithm is first introduced. Then, the details of the normal data extraction step and data marking step are separately explained.

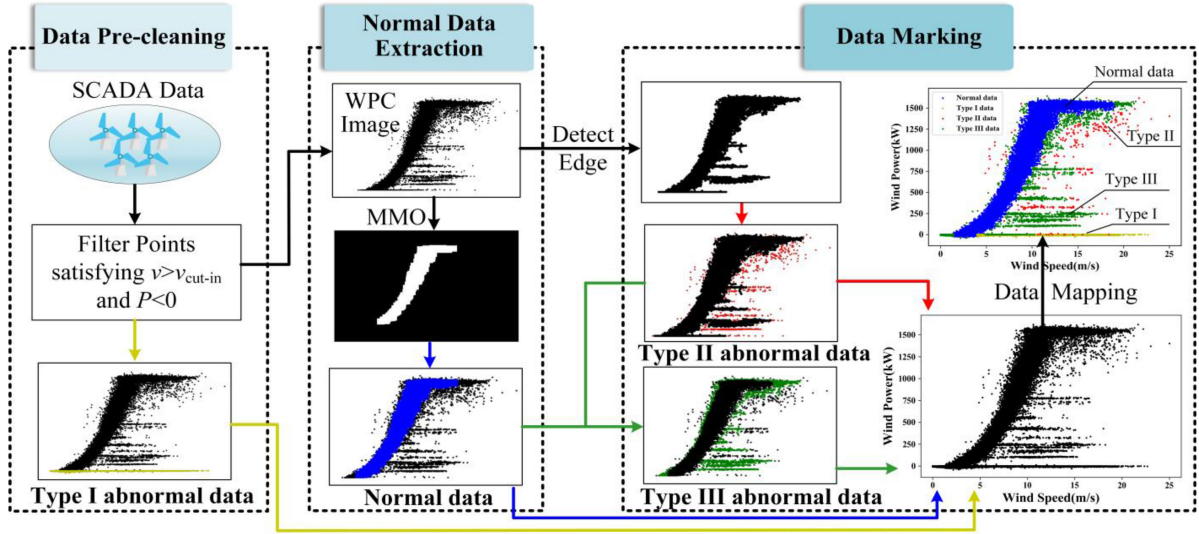

Fig. 2. Schematic diagram of image-based abnormal data detection and cleaning algorithm.

TABLE I
THE PROCEDURE OF IMAGE-BASED ABNORMAL DATA DETECTION AND CLEANING APPROACH

| | |
|---|---|
| **STEP 1: Pre-cleaning step** | |
| 1.1 | Filter and detect Type I abnormal points satisfying $v > v_{\text{cut-in}}$ and $P < 0$, where $v_{\text{cut-in}}$ is the cut-in speed. |
| **STEP 2: Normal data extraction** | |
| 2.1 | Generate the WPC binary image based on the scatter figure of the rest data; |
| 2.2 | Extract the principal part of the WPC binary image is by MMO under different sizes of the structuring element; |
| 2.3 | Calculate the Hu moments of the extracted principle part and reference WPC; |
| 2.4 | Determine the optimal size of the structuring element by minimizing the dissimilarity between the extracted principle part and reference WPC. The extracted principle part is represented as Normal data. |
| **STEP 3: Data marking** | |
| 3.1 | Recognize the image edge of the WPC image; |
| 3.2 | Filter the data located outside the image edge, identified as Type II abnormal data; |
| 3.3 | Filter the normal data and the rest part is regarded as Type III abnormal data; |
| 3.4 | Map the wind power points into binary image pixel points; |
| 3.5 | Mark the normal and three types of abnormal data based on the mapping relationship. |

### A. Image-Based Data Cleaning Algorithm Process

The schematic diagram of the imaged-based abnormal data detection and cleaning algorithm is displayed as Fig. 2.

Three steps, data pre-cleaning, normal data extraction and data marking, are included, shown as Table I. The process of the proposed algorithm is separated into three steps including data pre-cleaning, normal data extraction and data marking. Data pre-cleaning step filters the Type I abnormal points satisfying $v > v_{\text{cut-in}}$ and $P < 0$, where $v_{\text{cut-in}}$ is the cut-in speed. Normal data extraction step extracts the normal data part based on the WPC by the MMO and Hu moments. Data marking step first detects the Type II and Type III abnormal data in WPC by the extracted normal data part and edge detecting method, and then marks the SCADA points based on WPC detection results.

### B. Image-Based Normal Data Extraction

After filtering Type I abnormal data, the scatter figure of the rest data is converted to the WPC image. Based on gray value of each pixel point, the wind power curve greyscale image is further converted to the binary image. Then, the MMO [17], combined with the Hu moment [18], is used to extract the principle part of the WPC image. The Hu moment is employed to determine the optimal size of the structuring element in the mathematical morphology.

*1) Image Principle Part Extraction by MMO:* Mathematical morphology, developed from set theory, contributes a wide range of operators to binary image processing, such as the edge detection, noise removal, image enhancement and image segmentation [19], [20]. The basic operations, including erosion operation and dilation operation have two elements, the input binary image, donated as $A$, and the structuring element, donated as $B$. The structuring element consists of a pattern including an origin and some surrounding discrete points, shown as Fig. 3. The erosion operation diminishes the size of the object, which is employed to filter the scattered noise and the extra part of image, expressed by (1). The dilation operation increases the size of the object, which is utilized to filter the inner noise of the image and can be made directional by using less symmetrical structuring elements, expressed as (2).

$$A \ominus B = \bigcap_{b \in B} A_{-b} \quad (1)$$

$$A \oplus B = \bigcup_{b \in B} A_b \quad (2)$$

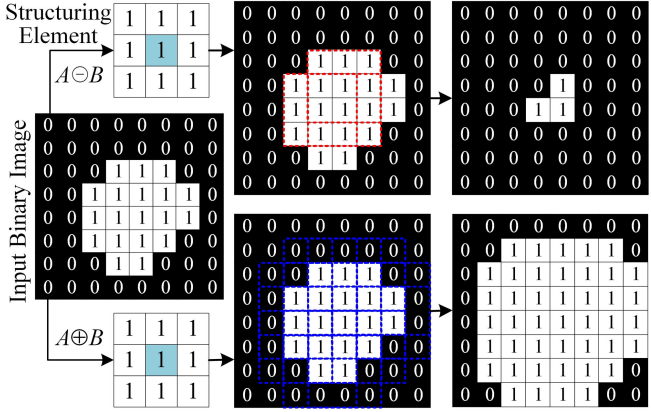

Fig. 3. Example of erosion and dilation operation by a 3 × 3 structuring elements.

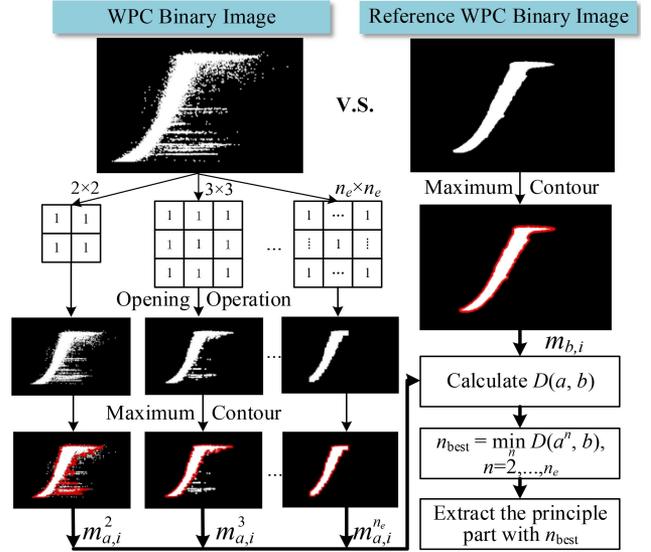

Fig. 4. The process of image-based normal data extraction.

where $\ominus$ and $\oplus$ separately denote the erosion and dilation operations, $b$ denotes the pixels of $B$.

The combinations of the erosion and dilation operations can generate other mathematical morphology operations. Opening operation, defined as $(A \ominus B) \oplus B$ is employed to identify the principal part of the wind power curve.

*2) Image-Based Dissimilarity Calculation by Hu Moments:* Due to the size of the structuring element, $B$, dramatically impacting the principle part extraction performance, the Hu Moment [18] is introduced to calculate the dissimilarity of the extracted principle part and the ideal power curve image. The Hu moment is the image invariant moment describing the image features which is not influenced by the image rotation, reflection, translation and scaling.

Suppose the wind power curve binary image $f(x, y)$ with $M \times N$ size, the $p + q$ order central moment of the image is defined as (3)

$$\mu_{pq} = \sum_{x=1}^{M} \sum_{y=1}^{N} (x - \bar{x})^p (y - \bar{y})^q f(x, y) \quad (3)$$

Where $\bar{x} = \sum\sum (xf(x,y))/\sum\sum f(x,y)$ and $\bar{y} = \sum\sum (yf(x,y))/\sum\sum f(x,y)$. In order to eliminate the impact of the image scaling, the central moment is normalized as (4).

$$\eta_{pq} = \mu_{pq}/\mu_{00}^{\gamma}, \quad \gamma = 1 + (p+q)/2, \, p+q = 2, 3, \ldots \quad (4)$$

The order two and three normalized central moments compose the seven invariant moments, $I_1, \ldots, I_7$, called as Hu moments [18]. The $I_1$ features the moment of inertia around the image's centroid reflecting the pixels' intensities, and $I_7$ is a skew orthogonal invariant useful in distinguishing mirror images.

In the process of image dissimilarity calculation, the Hu moments are further transferred as (5). The $m_{a,i}$ and $m_{b,i}$ are denoted as the transferred parameters of image $a$ and image $b$ by the $i$th Hu moments. Thus, the dissimilarity, $D(a, b)$, of the image $a$ and image $b$ is calculated as (6).

$$m_i = \text{sign}(I_i) \cdot \log(I_i), \quad i = 1 \ldots 7 \quad (5)$$

$$D(a, b) = \max_{i=1\ldots7} \frac{|m_{a,i} - m_{b,i}|}{|m_{a,i}|} \quad (6)$$

*3) The Process of Normal Data Extraction:* The process of the normal data extraction is displayed as Fig. 4. The reference WPC image, is introduced to evaluate the extracted principle part of the binary image, which comes from the normal wind turbine. The abnormal points of the normal wind turbine are first manually identified by human experts. The rest normal data generates the reference WPC image. With different size of the structuring element, $n$, the opening operation first works on the binary image to extract its principle part. To reduce the computational cost, the extracted principle part is replaced by its maximum contour [21]. Then the Hu moments are utilized to present the maximum contour of the binary image and reference power curve image. Based on Eqs. (5)-(6), the dissimilarity between the binary image and the reference one is calculated. The best size of the structuring element, $n_{\text{best}}$, is determined according to Eq. (7). The normal data of the wind power points are presented by the principle part extracted with $n_{\text{best}}$.

$$n_{\text{best}} = \min_{n} D(a^n, b), n = 2, \ldots, n_e \quad (7)$$

Where $n_e$ is the maximal size of the structuring element.

*C. Data Marking*

After detecting Type I abnormal data and the normal data, Type II and III abnormal data are identified sequentially. The abnormal data of the wind turbine data are marked by the mapping relationship.

In Fig. 1, the edge of the WPC image is first identified by the edge detecting method. The mutation of adjacent binary pixel points indicates the coordinates of the edge of the WPC image. The data, scattered outside the edge, are regarded as Type II abnormal data. The data, inside the detected edge and exclusive of the normal data, are regarded as Type III abnormal data.

TABLE II
THE WIND TURBINE SPECIFICATION OF TWO WIND FARMS

|  | Matang | Gaojiagou |
|---|---|---|
| Cut-in Speed(m/s) | 3.0 | 2.5 |
| Rated Speed(m/s) | 13.0 | 11.0 |
| Cut-out Speed(m/s) | 25.0 | 21.0 |
| Rate Power(kW) | 1500 | 1500 |
| No. of Wind Turbines | 17 | 20 |

To mark the actual wind turbine data based on the detection results of the WPC image, a data mapping algorithm is proposed to build the relationship between the wind power points and image pixel points.

Suppose the pixel point $(x, y)$ of WPC binary image $f(x, y)$, $x = 1, \ldots, M$, $y = 1, \ldots, N$, the $i$th wind power point $(v_i, P_i)$, the scaling parameters $(\Delta x, \Delta y)$ between the pixel point and wind power point is calculated as (8).

$$\Delta x = (x_{\max} - x_{\min})/(v_{\max} - v_{\min})$$
$$\Delta y = (y_{\max} - y_{\min})/(P_{\max} - P_{\min}) \qquad (8)$$

Where $x_{\max} = \max\{x \mid f(x, y) = 1\}$, $x_{\min} = \min\{x \mid f(x, y) = 1\}$, $y_{\max} = \max\{y \mid f(x, y) = 1\}$, $y_{\min} = \min\{y \mid f(x, y) = 1\}$, the $v_{\max}$ and $v_{\min}$ are the maximum and minimum wind speed, the $P_{\max}$ and $P_{\min}$ are the maximum and minimum wind power. The $i$th wind power point $(v_i, P_i)$ corresponding to the image pixel point $(x_i, y_i)$ is calculated as Eq. (9). Based on the data mapping algorithm, the normal and the abnormal wind power data from SCADA are marked.

$$x_i = x_{\min} + (v_i - v_{\min}) \times \Delta x$$
$$y_i = y_{\max} - (P_i - P_{\min}) \times \Delta y \qquad (9)$$

## IV. CASE STUDY

To validate the performance and generalization of the proposed algorithm, the wind turbine SCADA data from two wind farms, Matang wind farm, Jiangsu Province and Gaojiagou wind farm, Shanxi Province, China, are utilized. The CA [7] and LOF [6] algorithm are introduced as the comparison algorithms. The proposed algorithm and comparison algorithms are implemented by Python and operated on the laptop computer with CPU Inter(R) Core(TM) i7-5600 @2.60 GHz, RAM 8 GB.

The wind turbine specification of two wind farms is shown as Table II. The SCADA data of Matang wind farm are collected each 10 minutes from 2016/01/01 to 2016/08/31. The SCADA data of Gaojiagou wind farm are collected each 10 minutes from 2015/04/01 to 2016 /12/31.

### A. Performance of the Proposed Method

The image resolution of wind turbine WPC image is set $288 \times 432$ PPI (Pixels Per Inch). Each wind power point is represented by $2 \times 2$ PPI in the WPC image. In the reference

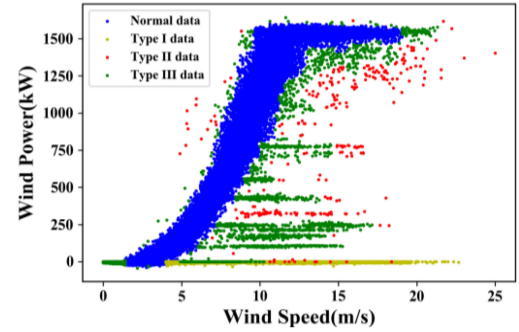

Fig. 5. The data detection and cleaning performance of the proposed on M-09.

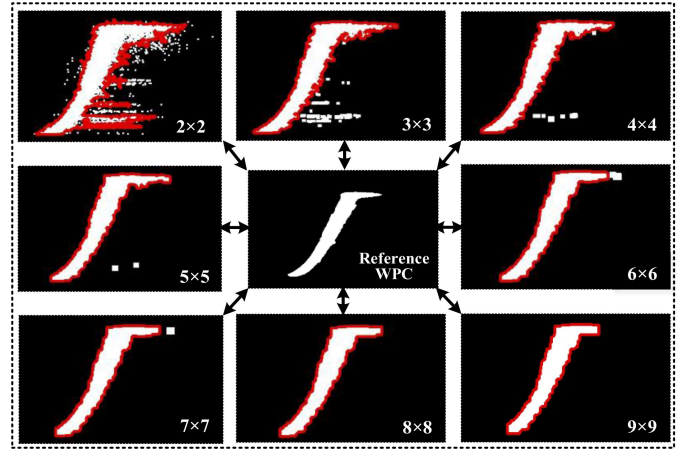

Fig. 6. The data detection and cleaning performance of the proposed on M-09.

TABLE III
THE DISSIMILARITY CALCULATION RESULT OF WIND TURBINE M-09 AND REFERENCE WPC

| n | D(a, b) | n | D(a, b) |
|---|---|---|---|
| 2 | 1.52761 | 3 | 0.380173 |
| 4 | 0.246093 | 5 | 0.217903 |
| 6 | 0.229587 | 7 | 0.226388 |
| 8 | 0.236105 | 9 | 0.285982 |

WPC, the min width of curve part is around 12 PPI. Thus, the candidate size of structuring element is set from $2 \times 2$ to $9 \times 9$.

The wind turbine M-09 in Matang wind farm is chosen as the example to show the performance of the proposed algorithm in detail. Fig. 5 presents the performance of the abnormal data detection and cleaning of M-09. The wind curtailment greatly impacts the shape of the wind power curve. It is obvious that the abnormal data are effectively detected and cleaned. Under different size of the structuring element, $n \times n$, the extracted principle part and corresponding maximum contour are shown as Fig. 6. It is seen that the area of the extracted normal part decreases with the increasing $n$. The dissimilarity between the maximum contours of the extracted principle part and the reference WPC is displayed in Table III. The dissimilarity grows down first and then goes up. It is implied that the excessive or too little information contained by the extracted principle part would influence the dissimilarity.

TABLE IV
THE ABNORMAL DATA CLEANING RESULT OF GAOJIAGOU WIND FARM

| WT | Number of data | Image-based detection and Cleaning algorithm | | | | | | | | LOF | | CA | | k-means | |
|---|---|---|---|---|---|---|---|---|---|---|---|---|---|---|---|
| | | Data Pre-cleaning | | Normal Data Extraction | | Data Marking | | Total | | | | | | | |
| | | R | T | R | T | R | T | R | T | R | T | R | T | R | T |
| G-01 | 31266 | 0.77 | 0.002 | 19.50 | 1.17 | 0 | 1.54 | 20.27 | 2.72 | 9.93 | 5.68 | 9.74 | 37.37 | 5.43 | 10.11 |
| G-02 | 31157 | 0.80 | 0.002 | 15.07 | 1.19 | 0 | 1.43 | 15.87 | 2.62 | 9.95 | 6.42 | 9.62 | 37.16 | 5.15 | 9.88 |
| G-03 | 31277 | 0.79 | 0.003 | 10.19 | 1.20 | 0 | 1.36 | 10.98 | 2.57 | 9.96 | 5.73 | 6.89 | 37.02 | 4.16 | 9.37 |
| G-04 | 30862 | 0.92 | 0.003 | 16.99 | 1.30 | 0 | 1.56 | 17.91 | 2.86 | 9.94 | 5.65 | 9.46 | 36.55 | 3.73 | 10.40 |
| G-05 | 30756 | 1.15 | 0.002 | 7.68 | 1.30 | 0 | 1.27 | 8.82 | 2.57 | 9.95 | 5.47 | 7.28 | 35.80 | 4.01 | 9.06 |
| G-06 | 30924 | 1.06 | 0.003 | 12.79 | 1.29 | 0 | 1.47 | 13.85 | 2.77 | 9.92 | 5.55 | 7.92 | 36.03 | 4.41 | 9.16 |
| G-07 | 31401 | 0.80 | 0.002 | 13.06 | 1.19 | 0 | 1.35 | 13.86 | 2.54 | 9.96 | 5.62 | 7.30 | 37.63 | 4.52 | 9.75 |
| G-08 | 31157 | 0.73 | 0.002 | 19.32 | 1.22 | 0 | 1.45 | 20.05 | 2.67 | 9.93 | 5.65 | 9.60 | 36.80 | 5.51 | 10.57 |
| G-09 | 31059 | 2.81 | 0.002 | 17.93 | 1.19 | 0 | 1.35 | 20.74 | 2.54 | 9.94 | 5.62 | 12.11 | 35.91 | 7.52 | 9.60 |
| G-10 | 31421 | 1.13 | 0.002 | 9.13 | 1.19 | 0 | 1.34 | 10.26 | 2.53 | 9.96 | 5.63 | 6.21 | 37.40 | 5.14 | 8.90 |
| G-11 | 31450 | 0.97 | 0.004 | 14.84 | 1.18 | 0 | 1.34 | 15.81 | 2.53 | 9.94 | 5.58 | 7.33 | 37.24 | 4.53 | 9.01 |
| G-12 | 31343 | 0.81 | 0.003 | 11.58 | 1.22 | 0 | 1.32 | 12.39 | 2.55 | 9.96 | 5.49 | 6.30 | 36.91 | 6.22 | 9.38 |
| G-13 | 23767 | 0.87 | 0.002 | 8.47 | 1.18 | 0 | 1.17 | 9.34 | 2.35 | 9.93 | 4.41 | 8.93 | 26.77 | 3.90 | 9.58 |
| G-14 | 31368 | 0.77 | 0.002 | 13.47 | 1.20 | 0 | 1.39 | 14.24 | 2.58 | 9.93 | 5.63 | 6.03 | 38.50 | 4.24 | 9.88 |
| G-15 | 31248 | 0.70 | 0.002 | 12.00 | 1.28 | 0 | 1.36 | 12.71 | 2.64 | 9.91 | 5.61 | 7.76 | 38.81 | 4.27 | 9.77 |
| G-16 | 31134 | 0.93 | 0.002 | 10.89 | 1.19 | 0 | 1.34 | 11.82 | 2.54 | 9.91 | 5.65 | 6.67 | 37.25 | 4.83 | 9.14 |
| G-17 | 31145 | 0.92 | 0.002 | 11.59 | 1.21 | 0 | 1.35 | 12.51 | 2.56 | 9.94 | 5.57 | 7.28 | 38.41 | 4.30 | 9.62 |
| G-18 | 29462 | 5.75 | 0.002 | 19.17 | 1.21 | 0 | 1.34 | 24.92 | 2.55 | 9.96 | 5.40 | 13.01 | 34.21 | 4.99 | 8.76 |
| G-19 | 31290 | 1.13 | 0.002 | 7.15 | 1.21 | 0 | 1.33 | 8.28 | 2.54 | 9.97 | 5.63 | 5.78 | 37.70 | 5.38 | 9.65 |
| G-20 | 30978 | 0.84 | 0.002 | 13.39 | 1.28 | 0 | 1.42 | 14.22 | 2.70 | 9.96 | 5.52 | 7.20 | 38.03 | 5.63 | 8.72 |

TABLE V
THE ABNORMAL DATA CLEANING RESULT OF MATANG WIND FARM

| WT | Number of data | Image-based detection and Cleaning algorithm | | | | | | | | LOF | | CA | | k-means | |
|---|---|---|---|---|---|---|---|---|---|---|---|---|---|---|---|
| | | Data Pre-cleaning | | Normal Data Extraction | | Data Marking | | Total | | | | | | | |
| | | R | T | R | T | R | T | R | T | R | T | R | T | R | T |
| M-01 | 93230 | 3.99 | 0.005 | 8.05 | 1.34 | 0 | 2.71 | 12.04 | 4.06 | 9.99 | 16.67 | 7.47 | 141.54 | 13.48 | 8.95 |
| M-02 | 90596 | 7.21 | 0.004 | 12.27 | 1.30 | 0 | 2.70 | 19.47 | 4.00 | 9.92 | 16.41 | 10.87 | 126.69 | 10.61 | 9.89 |
| M-03 | 93383 | 4.50 | 0.005 | 6.90 | 1.34 | 0 | 2.72 | 11.40 | 4.07 | 9.93 | 16.24 | 10.39 | 139.02 | 17.73 | 10.47 |
| M-04 | 93721 | 5.06 | 0.004 | 14.47 | 1.35 | 0 | 2.86 | 19.53 | 4.21 | 9.95 | 16.63 | 10.67 | 138.18 | 16.54 | 9.12 |
| M-05 | 93394 | 5.30 | 0.004 | 14.70 | 1.34 | 0 | 2.83 | 20.00 | 4.18 | 9.94 | 17.12 | 10.15 | 138.37 | 12.28 | 9.38 |
| M-06 | 92723 | 6.41 | 0.004 | 15.93 | 1.33 | 0 | 2.97 | 22.33 | 4.30 | 9.93 | 16.00 | 12.79 | 138.14 | 10.94 | 9.20 |
| M-07 | 93702 | 5.88 | 0.003 | 13.25 | 1.31 | 0 | 2.89 | 19.13 | 4.21 | 9.93 | 16.77 | 11.32 | 141.54 | 10.92 | 10.26 |
| M-08 | 92095 | 3.35 | 0.003 | 12.94 | 1.33 | 0 | 2.80 | 16.29 | 4.14 | 9.91 | 17.51 | 7.21 | 132.54 | 11.76 | 10.66 |
| M-09 | 92937 | 7.66 | 0.004 | 11.19 | 1.35 | 0 | 2.80 | 18.85 | 4.16 | 9.89 | 16.43 | 13.32 | 137.75 | 11.49 | 9.24 |
| M-10 | 92818 | 5.18 | 0.004 | 8.60 | 1.33 | 0 | 2.72 | 13.78 | 4.06 | 9.94 | 16.27 | 8.94 | 140.64 | 11.74 | 9.72 |
| M-11 | 94218 | 3.74 | 0.004 | 5.19 | 1.35 | 0 | 2.77 | 8.93 | 4.12 | 9.94 | 18.88 | 9.38 | 159.67 | 13.63 | 10.50 |
| M-12 | 94384 | 2.61 | 0.003 | 9.63 | 1.35 | 0 | 2.86 | 12.24 | 4.21 | 9.94 | 17.18 | 5.12 | 158.33 | 16.22 | 9.95 |
| M-13 | 86437 | 4.31 | 0.005 | 12.18 | 1.32 | 0 | 2.72 | 16.49 | 4.04 | 9.94 | 15.08 | 6.81 | 142.98 | 10.17 | 10.13 |
| M-14 | 88006 | 5.92 | 0.004 | 12.25 | 1.34 | 0 | 2.77 | 18.17 | 4.11 | 9.94 | 15.47 | 7.48 | 143.52 | 12.36 | 10.13 |
| M-15 | 87084 | 3.89 | 0.004 | 10.36 | 1.34 | 0 | 2.72 | 14.25 | 4.06 | 9.95 | 15.47 | 7.10 | 143.22 | 11.40 | 10.12 |
| M-16 | 87907 | 4.11 | 0.004 | 9.15 | 1.33 | 0 | 2.68 | 13.26 | 4.01 | 9.89 | 15.77 | 5.61 | 139.22 | 14.81 | 10.05 |
| M-17 | 87198 | 5.99 | 0.003 | 10.79 | 1.36 | 0 | 2.63 | 16.78 | 3.99 | 9.95 | 15.99 | 5.12 | 135.51 | 14.12 | 9.40 |

The data cleaning results of all wind turbines in two wind farms are shown as Table IV and Table V, including the abnormal data deletion rate, $R(\%)$, and computational time, $T(s)$. The algorithm settings of the rest turbines and the reference WPC are the same as the M-09 to prove the universality of the proposed method. The computational time of data pre-cleaning and normal data extraction is almost not impacted by the data amount. The time of the data marking step grows up with the amount of the data. Based on the pre-cleaning data deletion rate, the wind curtailment of Matang wind farm appears more frequently than Gaojiagou wind farm.

B. Comparative Experiments

To verify the effectiveness of the proposed algorithm, the LOF method, CA method and k-means method, are introduced.

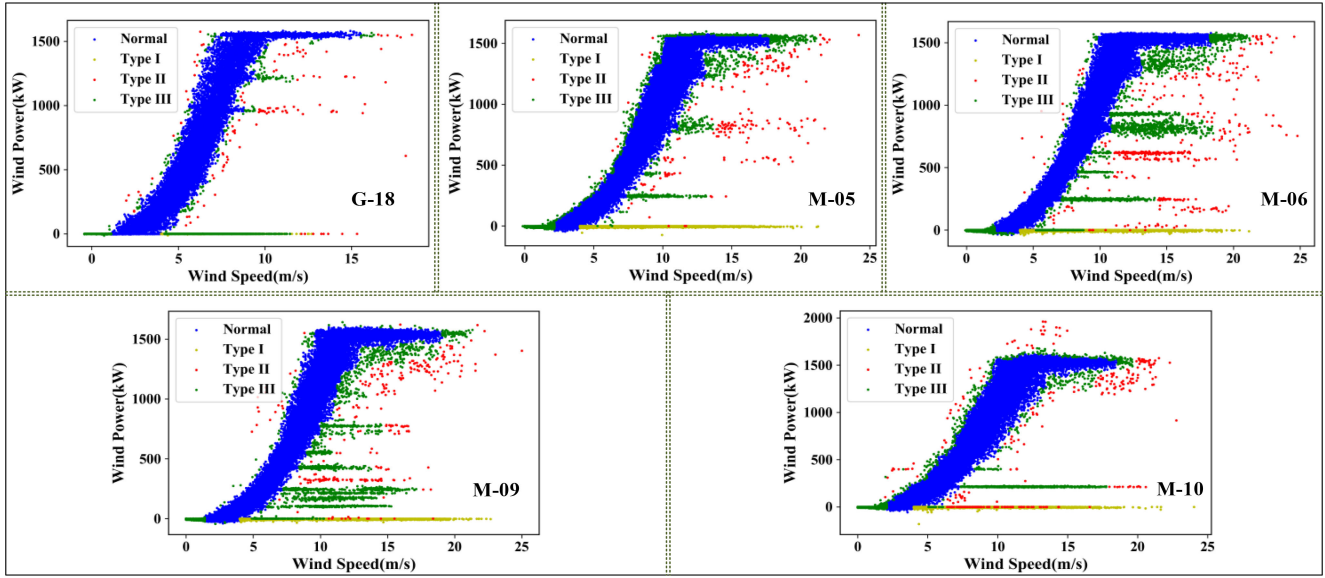

Fig. 7. The data cleaning result by the proposed image-based algorithm.

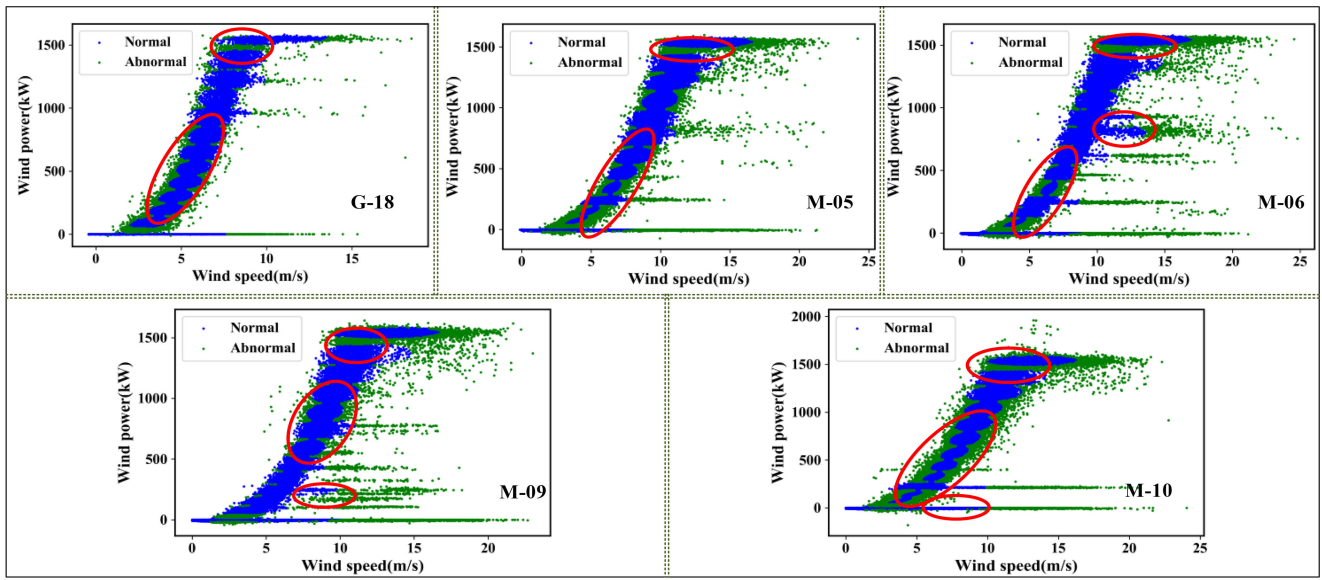

Fig. 8. The data cleaning result by $k$-means method.

The LOF algorithm is a typical clustering algorithm to filter the outliers. The LOF score of each point is calculated based on the weighted distance of $k$ nearest neighbor points to evaluate its isolation level. The outliers are identified by comparing their LOF score with the LOF threshold. In our experiments, the number of neighboring points, $k$, is set to 300 and the LOF threshold is set to 10% based on the performances of the other three method in the experiments. In CA method, the point algorithm first cleans part of abnormal data through detecting the change point of variance change rate. The grouping-quartile algorithm then extracts the normal data from the rest data. The parameter, wind speed interval, is selected as 0.5 m/s by convention [7]. The $k$-means clustering algorithm is a typical data-driven method to clean the abnormal data and has been utilized in the previous reference [8]. The number of clusters is set as 13 according to the reference article. The data cleaning results of LOF, CA and $k$-means, are displayed in Tables IV–V.

The $R$ of LOF is close to 10% in all cases due to the setting of the LOF threshold. It is concluded that the LOF threshold directly determines the algorithm performance. The selection of the LOF threshold also brings the difficulty and makes LOF lack the flexibility. The abnormal data rate of the proposed algorithm and CA varies case by case, which implies their better adaptive capacity compared with LOF. The $R$ of $k$-means differs in the two wind farms a lot. The performance of $k$-means is data-dependent and the number of clusters should be set case by case to obtain the best result. Besides, the proposed algorithm costs less time than the LOF, CA algorithms and $k$-means. The CA algorithm costs the most time because of the variance change rate calculation of each wind speed interval.

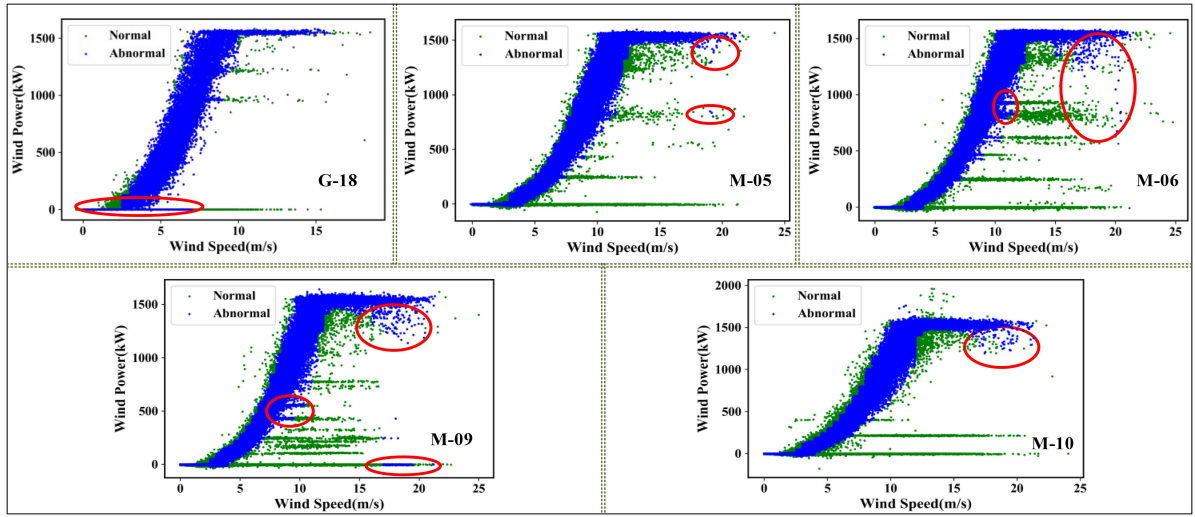

Fig. 9. The data cleaning result by CA method.

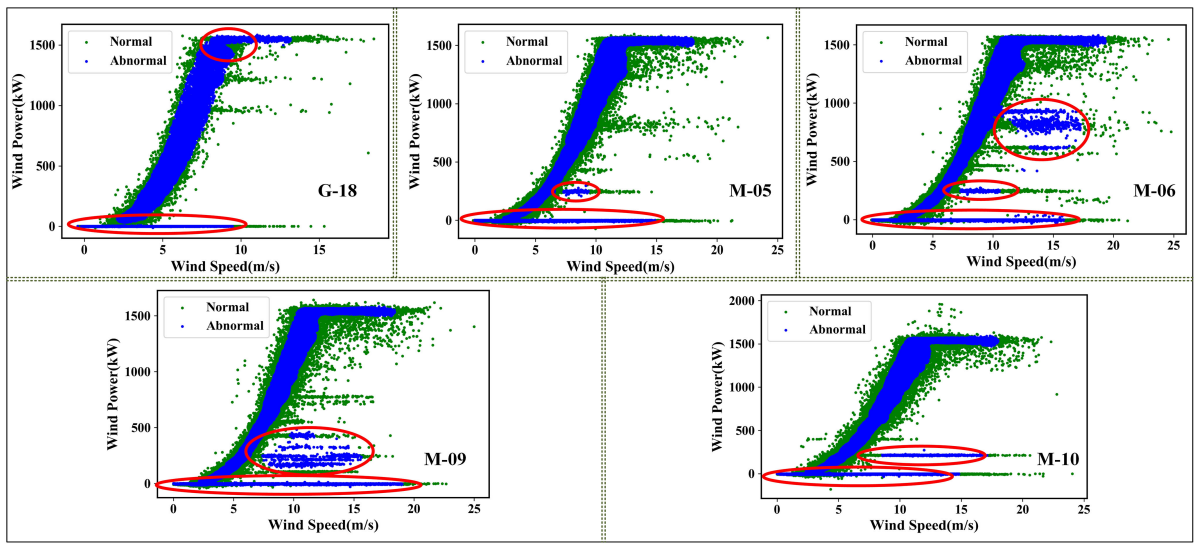

Fig. 10. The data cleaning result by LOF method.

All the results are achieved under the same setting and reference WPC, it can be proved the image-based algorithm is robust to the distribution and amount of abnormal data and the reference WPC. The average $R$ of the image-based algorithm is more than LOF, CA and $k$-means. It is because the proposed method can detect more abnormal data caused by the too small wind speed or the wind curtailment command. The $R$ of wind turbines, G18, M-05, M-06, M-09 and M-10, detected by the image-based method, is much greater than LOF, CA and $k$-means, selected as the examples and presented in Fig. 7 Fig. 10.

In Figs. 7–10, it is obvious the image-based method outperforms LOF, CA and $k$-means, especially in the case of tremendous stacked abnormal data. The cleaning performance of LOF is the worst. In Figs. 7–10, the wind curtailment of M-06, M-09 and M-10, is serious. LOF, CA and $k$-means do not completely filter the abnormal data due to the impacted LOF score and the variance change rate. The image-based algorithm utilizes the image instead of the actual data to undermine the influence of the abnormal data density. Thus, it can achieve the good performance when abundant abnormal data exists and bring in the high abnormal data rate. Besides, the image-based algorithm can also clearly provide the category information of the abnormal data.

Compared with the LOF, CA and $k$-means, the proposed image-based algorithm has several advantages: i) It is more intuitive due to it directly working on the WPC image; ii) since the proposed algorithm can obtain the good performance for different wind turbines with the same parameter settings, it has strong generalization ability; iii) the computational time of the proposed method is much less than LOF, CA and $k$-means, especially for the large amount of data; iv) it can clearly provide the additional and valuable category information of the abnormal data for wind farm operators. Due to its generalization, computational efficiency and effectiveness, the actual application potential of the proposed algorithm is much greater than LOF, CA and $k$-means.

## V. Conclusion

This paper proposed an image-based abnormal data detection and cleaning algorithm for wind power data. The wind power curve data were converted to the binary image, the principle part of the image, representing the normal data, was extracted by the MOO and Hu moments. Minimizing the dissimilarity between the extracted principle part and the reference wind power curve was to determine the optimal size of the structuring element. The LOF, CA and $k$-means algorithm were introduced as the comparison approaches. The 33 wind turbines data from two wind farms validated the efficiency and effectiveness of the proposed image-based algorithm. Due to the shape of the wind power curve deciding the performance rather than the data density, the proposed algorithm has greater capacity to filter the stacked abnormal data than LOF, CA and $k$-means. The good abnormal data filtering performance were obtained for all wind turbines from different wind farms under the same parameter setting and reference WPC. It was proved the generalization of the proposed algorithm, which could promote its actual application. Besides, the proposed algorithm could not only perform well in filtering the abnormal data, but also provide their category information for wind farm operators.

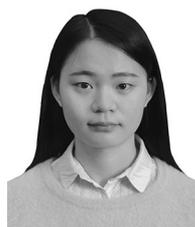

**Huan Long** (M'15) received the B.Eng. degree from the Department of Automation, Huazhong University of Science and Technology, Wuhan, China, in 2013, and the Ph.D. degree in systems engineering and engineering management from the City University of Hong Kong, Hong Kong, in 2017. She is an Assistant Professor with the School of Electrical Engineering, Southeast University, Nanjing, China. Her research fields include data mining applied in modeling, optimizing, monitoring the renewable energy system and power system.

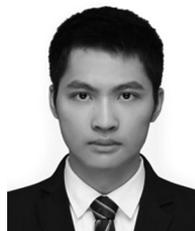

**Linwei Sang** received the B.Eng. degree from the School of Electric Engineering, Southeast University, Nanjing, China, in 2018. He is currently working toward the master's degree in the electric engineering, Southeast University, Nanjing, China. His research includes machine learning, data mining, and the control of the distributed energy, such as the cleaning of abnormal data in wind power curve, load prediction by artificial intelligence.

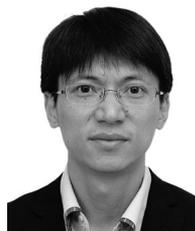

**Zaijun Wu** (M'12) received the B.Eng. degree in power system and its automation from the Hefei University of Technology, Hefei, China, in 1996, and the Ph.D. degree in electrical engineering from Southeast University, Nanjing, China, in 2004. He was a Visiting Scholar with the Ohio State University, USA, from 2012 to 2013. He is currently a Professor with the School of Electrical Engineering, Southeast University, Nanjing, China. His research interests include micro-grid, active distribution network, and power quality. He is the author or co-author of more than 100 referred journal papers, and a reviewer of several journals.

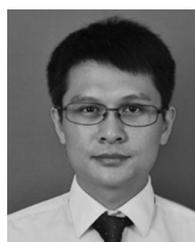

**Wei Gu** (M'09–SM'16) received the B.S. and Ph.D. degrees in electrical engineering from Southeast University, Nanjing, China, in 2001 and 2006, respectively. From 2009 to 2010, he was a Visiting Scholar with the Department of Electrical Engineering, Arizona State University, Tempe, AZ, USA. He is currently a Professor with the School of Electrical Engineering, Southeast University. He is also the Director of the Institute of Distributed Generations and Active Distribution Networks. His research interests include distributed generations and microgrids, and integrated energy system.